\documentclass[aps,pra,showpacs,superscriptaddress,floatfix,onecolumn,preprint]{revtex4-1}

\usepackage{graphicx}
\usepackage{dcolumn}
\usepackage{bm}
\usepackage{setspace}
\usepackage{psfrag}
\usepackage{subfigure}
\usepackage{amsfonts}
\usepackage{braket}
\usepackage{color}

\newcommand{\be}{\begin{equation}}
\newcommand{\ee}{\end{equation}}
\newcommand{\bea}{\begin{eqnarray}}
\newcommand{\eea}{\end{eqnarray}}

\newcommand{\bef}{\begin{figure}}
\newcommand{\enf}{\end{figure}}

\begin{document}

\title{Nonadiabatic tunneling in circularly polarized laser fields.\\ III. Comparison with experiment}

\author{Ingo Barth}
%\address{Max Born Institute, Max-Born-Str. 2A, 12489, Berlin, Germany}
%\ead{barth@mbi-berlin.de}
\author{Olga Smirnova}
\affiliation{Max Born Institute, Max-Born-Str. 2A, 12489, Berlin, Germany}
%\ead{smirnova@mbi-berlin.de}
\date{\today}

\begin{abstract}
We  compare results of the recent experiment by Herath $et~al$ [T. Herath, L. Yan, S. K. Lee, and W. Li, Phys. Rev. Lett. \textbf{109}, 043004 (2012)]
on strong-field nonadiabatic tunneling in circularly polarized laser fields with the original predictions of our theory
 [I. Barth and O. Smirnova, Phys. Rev A \textbf{84}, 063415 (2011)] that stimulated these experiments.
 We show that the theory and experiment are in very good agreement. We also explain why the initial comparison performed by Herath $et~al$
% [T. Herath, L. Yan, S. K. Lee, and W. Li, Phys. Rev. Lett. \textbf{109}, 043004 (2012)]
 has suggested quantitative discrepancies with our theory. We confirm that these seeming discrepancies are removed with an accurate application of our theoretical model.
We suggest an experiment for unique determination of the ionization preference of valence orbitals $p_+$ or $p_-$.
  \end{abstract}
\pacs{42.50.Hz, 32.80.Rm, 33.80.Wz }
%\noindent{\it Keywords}: strong field ionization, tunneling, ring currents, circular dichroism
\maketitle

%Tunneling perspective on strong field ionization \cite{Keldysh} is one of the most beautiful and productive in strong field physics.
%Strong low-frequency laser field creates  the tunneling barrier for the electron.
Strong field ionization can be understood as electron tunneling through the barrier created by the laser field and the core potential.
Sensitivity of the tunneling rate in circularly polarized fields to the sense of electron rotation in the initial state \cite{PRA1,PRA2} is a purely non-adiabatic effect,
associated with the rotation of the barrier. The kinematics of the electron motion  in classically
forbidden region is responsible for higher ionization rate for an electron counter-rotating with respect to the laser field prior to ionization \cite{PRA1,PRA2}.
For example,  right circularly polarized laser field preferentially removes an electron from the valence $p_-$ rather than $p_+$ orbital, creating
a counter-rotating hole in the system.
This asymmetry has recently been experimentally observed  in an elegant experiment of Herath $et~al$ \cite{herath}.
Asymmetry in ionization from  $p_-$ and $p_+$ orbitals  opens unique opportunities for
production of intense ultrashort bursts of spin-polarized electrons (and ions) in strong-field ionization \cite{spinpol}.

The experiment of Herath $et~al$ \cite{herath} used two time-delayed (500\,fs) circularly polarized laser pulses
with same (RR) and opposite (LR) polarizations to induce double ionization of Ar atom.
By monitoring the yield of sequential double ionization (SDI) triggered by the RR or LR sequence
of pulses, one can infer the sensitivity of ionization to the pulse helicity in each step.
Indeed, as suggested in \cite{herath}, the ratio of the SDI yields recorded in each experiment $\frac{I_\mathrm{SDI-LR}}{I_\mathrm{SDI-RR}}$ can be related to the
ratios  of the ionzation rates $\frac{w_+^{p_-}}{w_+^{p_+}}$ from $p_m$ orbitals of Ar ($\alpha$)and Ar$^+$ ($\alpha'$ ) in right circularly polarized fields  as follows:
\begin{eqnarray}
\label{SDI}
\frac{I_\mathrm{SDI-LR}}{I_\mathrm{SDI-RR}}=\frac{\alpha\alpha'+1}{\alpha+\alpha'}.
\end{eqnarray}
Performing a sequence of measurements in their ingenious set-up, Herath $et~al$ managed to reconstruct $\alpha$ and $\alpha'$.
However, the comparison with the theory of Ref. \cite{PRA1} performed in Ref.\cite{herath} suggested a discrepancy,
which we address below.

Time-delayed linearly polarized pump and probe laser pulses used in the experiment  are reported \cite{herath} to have
the following mean intensities  $\bar I_\mathrm{lin}=9\times 10^{13}$\,W/cm$^2$ and
$\bar I_\mathrm{lin}'=1.4\times 10^{14}$\,W/cm$^2$.
These values correspond to the amplitudes of the electric fields $\mathcal{E}_\mathrm{lin}=0.051$\,a.u.\ and $\mathcal{E}_\mathrm{lin}'=0.063$\,a.u..
Herath $et~al$ used a quarter wave plate to produce circularly polarized light, where the angle between the polarization of the
incident wave and the optical axis of the crystal is adjusted to $45^\circ$.
After the waveplate, the energy in the pulse stays the same but the amplitudes of the parallel and perpendicular components of the circularly polarized
electric field vector have reduced peak values compared to linear polarization,
$\mathcal{E}_x=\mathcal{E}_y=\mathcal{E}_\mathrm{lin}\cos(45^\circ)=\mathcal{E}_\mathrm{lin}/\sqrt{2}$. The same
applies to the amplitude of the circularly polarized electric laser field, i.e.\
$\mathcal{E}_\mathrm{circ}=\mathcal{E}_\mathrm{lin}/\sqrt{2}$. The corresponding values for circularly polarized
pump and probe laser pulses are $\mathcal{E}_\mathrm{circ}=0.036$\,a.u.\ and $\mathcal{E}_\mathrm{circ}'=0.045$\,a.u.\
The intensities of the circularly polarized pulses are $I_\mathrm{circ}=\bar I_\mathrm{lin}$ and $I'_\mathrm{circ}=\bar I'_\mathrm{lin}$.

Using these values for the amplitudes $\mathcal{E}_\mathrm{circ}$ and $\mathcal{E}'_\mathrm{circ}$ and the laser frequency $\omega=0.057$\,a.u.\ (800\,nm), we get
the Keldysh parameters $\gamma=1.7$ for Ar and $\gamma'=1.8$ for Ar$^+$. Applying analytical formulas (6)--(9) from \cite{PRA1},
 we obtain the ratios of the ionization rates $\alpha=5.8$ and $\alpha'=6.2$, and the SDI ratio $3.1$ in good agreement with
 the experimental result $3.6$ before the focal averaging is taken into account.
The ratios $\alpha=5.8$ and $\alpha'=6.2$  agree with the the experimental value $\alpha=\alpha'=7.1$ within the error-bars established in the experiment \cite{herath}.

Interestingly, in contrast to the common intuition, also shared by Herath $et~al$,  that the intensity and the focal volume averaging generally tend to
reduce observed  strong-field effects,  the asymmetry discussed here is an example of a  strong field effect that {\it benefits} from focal and intensity averaging.
Indeed, we have shown in  Refs.\,\cite{PRA1,PRA2}  that the ratio of the ionization rates  depends on the laser intensity and increases as the intensity decreases.
Increasing asymmetry of the ionization rates for larger values of the Keldysh parameter is a direct consequence of the non-adiabatic nature of the effect \cite{PRA1,PRA2}.
Thus, lower intensities at the wings of the pulse yield smaller ionization rates but larger ratios
$\alpha$ and $\alpha'$. Consequently, the laser intensity and the focal volume averaging can only enhance the ratio of the ionization rates.
To illustrate this remark, we consider the spatio-temporal intensity profile of the Gaussian beam
\begin{eqnarray}
I(r,z,t)&\sim&\left(\frac{w_0}{w(z)}\right)^2e^{-\frac{2r^2}{w(z)^2}-\frac{t^2}{2\sigma_t^2}}
\end{eqnarray}
where $w_0$ is the waist size, $w(z)=w_0\sqrt{1+z^2/z_R^2}$ is the $z$-dependent spot size, $z_R=\pi w_0^2/\lambda$ is the Rayleigh range, and $\sigma_t$ is the temporal Gaussian width.
Evaluating the corresponding averaged ratio
\begin{eqnarray}
\bar \alpha=\frac{\int_0^\infty r\,dr \int_{-\infty}^\infty dz\int_{-\infty}^\infty w^{p_-}_+(I(r,z,t))\,dt}{\int_0^\infty r\,dr \int_{-\infty}^\infty dz \int_{-\infty}^\infty w^{p_+}_+(I(r,z,t))\,dt},\label{aver}
\end{eqnarray}
that is independent of $w_0$, $z_R$ and $\sigma_t$,
we obtain $\bar\alpha=6.6$ and $\bar\alpha'=6.7$ from Eq. (\ref{aver}). If only temporal averaging is performed, $\bar\alpha=6.0$ and $\bar\alpha'=6.3$.
As we have anticipated, both averaged values are larger than the peak values obtained above, $\alpha=5.8$ and $\alpha'=6.2$. The averaged values are even
closer to the experimentally estimated number  $\alpha=\alpha'=7.1$ and yield the theoretical result of the SDI ratio $3.4$, in a very good
agreement  with the experimental result $3.6$.
Thus, the experimental results \cite{herath} are in good agreement with the theoretical predictions \cite{PRA1,PRA2}.

Unfortunately, the theoretical predictions of \cite{PRA1,PRA2} could not be tested in this experiment fully. Namely, the experiment has confirmed
the existence of the asymmetry but the detection scheme could not establish the sign of the effect -- the ionization preference of $p_-$ vs $p_+$ or vice versa
\cite{herath}. Herath $et~al$ have suggested an interesting  experiment utilizing an  oriented atomic sample with known helicity. Such sample
can be produced by photodissociation on the excited molecular state of a molecule such as ICl \cite{ICl}.
Here, we suggest another experiment for determining the value of $\alpha$, using the spin-orbit splitting.
Using the Stern-Gerlach-type experiment, four sub-levels $M_\frac32=\pm \frac12, \pm \frac32$  of a neutral iodine atoms in the electronic ground state $^2\!P_{\frac32}$
can be spatially separated and addressed separately \cite{Ulli}.
Measuring only the ratio of the ionization rates $\alpha$ for each sublevel of the state $^2\!P_{\frac32}$, one can extract the ionization rates for the orbitals $p_m$
%Since the original ratios of the ionization rates for atomic orbitals $p_\pm$ in the electronic ground state are equal to the ones for the spin-orbitals $p_{\frac32\pm\frac32}$ \cite{spinpol}, one can
and determine the ionization preference of the valence orbitals $p_+$ or $p_-$ in circularly polarized laser fields.
%Using the ionization potential $I_p=10.45$\,eV for iodine in the ionization channel from $^2\!P_\frac32$ to $^3\!P_2$,
%we estimate the theoretical ratio of the ionization rates between $p_-$ and $p_+$ valence orbitals
%as $\alpha=$.

In summary, we have compared theoretical and experimental results for ratios of the ionization rates from valence orbitals $p_\pm$ of Ar and Ar$^+$ in
circularly polarized laser fields and have found that they are in a very good quantitative agreement, resolving the seeming discrepancy in the
literature \cite{herath}.

We gratefully acknowledge the  support of the DFG (Grant No. Sm 292/2-1).
We thank Prof. U. Eichmann for useful and fruitful discussions.

\end{document}